\documentclass[twoside,fleqn]{article}
\usepackage{npb}
\usepackage{graphics}
\newcommand{\stackM}{\stackrel{\scriptstyle >}{{ }_{\sim}}}

\newcommand{\gsim}{\stackrel{\scriptstyle >}{{ }_{\sim}}}

\newcommand{\tbH}{\ensuremath{H^+\bar{t}b}}
\newcommand{\gb}{\ensuremath{\bar{b}g}}

\newcommand{\mH}{\ensuremath{M_{H^+}}}
\newcommand{\mb}{\ensuremath{m_b}}

\newcommand{\mt}{\ensuremath{m_t}}

\newcommand{\pptbH}{\ensuremath{p\bar{p}(pp)\to\tbH +X}}

\newcommand{\Dmb}[1][]{\ensuremath{\Delta\mb^{#1}}}
\newcommand{\aS}[1][]{\ensuremath{\alpha_S^{#1}}}

\newcommand{\tb}[1][]{\ensuremath{\tan^{#1}\!\beta}}

\newcommand{\msb}[1]{\ensuremath{m_{\tilde{b}_{#1}}}}
\newcommand{\mst}[1]{\ensuremath{m_{\tilde{t}_{#1}}}}
\newcommand{\be}{\begin{equation}}
\newcommand{\ee}{\end{equation}}

\newcommand{\GeV}{\ensuremath{\,{\rm GeV}}}

\title{Supersymmetric effects on heavy charged Higgs boson production in hadron colliders
\thanks{Talk presented at the \textit{6th International Symposium on
  Radiative Corrections, Application of
Quantum Field Theory to Phenomenology (RADCOR 2002), and the 6th
Zeuthen Workshop on Elementary Particle Theory, Loops and Legs in
Quantum Field Theory}, Kloster Banz, Germany, 8-13 September,
2002.}}

\author{Alexander Belyaev \address{Physics Department, Florida State University
Tallahassee, FL 32306-4350, USA}, Jaume Guasch \address{Theory
Group LTP, Paul
    Scherrer Institut, CH-5232 Villigen PSI, Switzerland.},
 \underline{Joan Sol\`a} \address{Departament d'Estructura i Constituents de la
Mat\`eria,  Universitat de Barcelona, Diagonal 647, E-08028 Barcelona,
 Catalonia, Spain,
and
Institut de F\'{\i}sica d'Altes Energies,
Universitat Aut\`onoma de
  Barcelona, E-08193 Bellaterra., Barcelona, Catalonia, Spain.}%
}

\begin{document}

\begin{abstract}
The production of a heavy supersymmetric charged Higgs boson
($M_{H^{\pm}}\ge 200\,GeV$) at the Tevatron and at the LHC is
studied. We include the leading one-loop quantum effects within
the MSSM in the relevant high $\tb$ region. Whereas the chances
for the Tevatron are limited, and critically depend on the size
of the unknown NLO QCD effects, at the LHC the discovery range is
more comfortable and may extend the reach above $M_{H^{\pm}}=
1\,TeV$.
\begin{flushright}
{{UB-ECM-PF-02-22,
hep-ph/0210XXX }}
\end{flushright}
\end{abstract}

\maketitle

\section{Introduction}
The experimental accuracy reached by the electroweak precision
observables in particle accelerators allows to test the Standard
Model (SM) at the quantum level within an accuracy better than
one per mil. Although precision observables at LEP have played an
important role in this respect, a new generation of experiments
at the Tevatron (II), LHC and a future Linear Collider (LC) will
be necessary to find out the ultimate nature of the spontaneous
symmetry breaking mechanism.  When extending the Higgs sector, we
mainly focus on two-Higgs-doublet models (2HDM), and most
particularly on the Type II ones --like that of the the Minimal
Supersymmetric Standard Model (MSSM)\,\cite{Haber}. The fact that
the Higgs bosons of the MSSM (and their SUSY counterparts) couple
most preferentially to heavy quarks suggests that the physics of
the top and bottom quarks is the natural arena where to try to
get a hint of the existence of SUSY Higgs particles. In this
paper we report on the production of a heavy charged Higgs boson
in association with the top and bottom quarks at the Tevatron and
LHC energies within the MSSM. The mass of the charged Higgs is
assumed $M_{H^{\pm}}\stackM 200\,GeV$, and so it cannot come from
the decay of a top quark. While the detection of a charged Higgs
boson would still leave a lot of questions unanswered, it would
immediately offer (in contrast to the detection of a neutral one)
indisputable evidence of physics beyond the SM. The basic
production processes under consideration are
\begin{equation}
\pptbH\hspace{1cm}\mathrm{Tevatron\;(LHC)}\,.  \label{tbh}
\end{equation}
There are many other processes for charged Higgs production,
which have been studied in the literature\,\cite{HiggsRunII},
such as: i) one-loop gluon fusion ($g\,g\rightarrow
H^{+}\,H^{-}$); ii) tree-level pair production in bottom quark
scattering ($b\,\bar{b}\rightarrow H^{+}\,H^{-}$), iii) associated
production with $W^{\pm}$: $b\bar{b}\rightarrow W^{\pm}\,H^{\mp}$
(tree-level)\,, $g\,g\rightarrow W^{\pm}\,H^{\mp}$ (one-loop), iv)
Drell-Yan type production ($q\bar{q}\rightarrow V\rightarrow
H^{+}\,H^{-}$). However, in all these cases the production
cross-section is smaller than for (\ref{tbh}) if we restrict
ourselves to high values of $\tb$  ($>20$). Although process
(\ref{tbh}) has been studied previously\,\cite{HiggsRunII,Hppt},
the inclusion of the leading quantum SUSY effects was missing and
turns out to be very important, as first hinted
in\,\cite{Coarasa} and further elaborated in\,\cite{bggs}.

\section{Leading order Cross-section in QCD}

At the parton level, the reaction (\ref{tbh}) proceeds through
three channels: i) $q\bar{q}$-annihilation for light quarks
\begin{equation}
q\bar{q}\rightarrow \tbH,  \label{qq-tbh}
\end{equation}
where $q=u,d$ (the $s$ contribution can be safely neglected), a
channel only relevant for the Tevatron; ii) $gg$-fusion
\begin{equation}
gg\rightarrow \tbH\,,  \label{gg-tbh}
\end{equation}
which is dominant at the LHC, but it can also be important at the
Tevatron for increasing $H^{+}$ masses ;
and finally there is the iii) bottom-gluon 2-body channel
\begin{equation}
\bar{b}g\rightarrow H^{+}\bar{t}\,\,.  \label{gb-th}
\end{equation}
We will compute the cross-section for the charged Higgs boson production process (%
{\ref{tbh}}) at the leading order (LO) in QCD, namely at
  $\mathcal{O}(\alpha
_{S}^{2})$.  However, the QCD corrections at the next-to-leading
order (NLO) or $\mathcal{O}(\alpha _{S}^{3})$ could be important.
Although a dedicated calculation of these higher order effects for
the process under consideration is not available, let us mention
that the corresponding calculation
for the related process in the Standard Model, $p\bar{p}%
(pp)\rightarrow H\,t\bar{t}+X$, is available and shows that the
NLO effects lead to a non-trivial QCD $K$-factor which is
typically smaller than one at the Tevatron, and up to $1.4$ at the
LHC\,\cite{Reina}.  On the other hand, partial calculations of
the QCD corrections to the process (\ref{tbh})  at the NLO have
appeared. For example, in Ref.~\cite{shouhua} the standard NLO QCD
corrections to the subprocess~(\ref{gb-th}) at the LHC are
computed, obtaining a large K-factor between $\sim1.6$ and $\sim
1.8$ for $\tb\gsim 20$. Taking into account these considerations,
a large $K$-factor for the full process (\ref{tbh}) coming from
the standard QCD corrections
 (e.g. $K^{\rm QCD}\simeq 1.5$) is not ruled out.  However, in the
 absence of a complete calculation of these QCD effects, in
 what follows we present our results at the leading order and will
 parametrize our ignorance of the complete NLO effects in terms of a $K$-factor, which will
 be assumed different from one only for the Tevatron,
 where the size of the cross-section is hardly the necessary for a
 viable study.  On the other hand, for the LHC the cross-section,
 being typically three orders of magnitude larger, can afford a QCD
 $K$-factor of one.

 Once a PDF for $b$-quarks is used, there is some amount of overlap between $%
\gb$- and $gg$-initiated amplitudes, which has to be removed~\cite
{Barnett}. The overlap arises because the $b$-density in the
$\gb$ amplitude receives contributions from gluon splitting which
was already counted in the $gg$ amplitude, so we have to avoid
double counting by  the subtracting of the gluon splitting term.
The net partonic cross-section from the $\gb$- and $gg$-initiated
subprocesses is
\begin{eqnarray}
& \sigma (\gb+gg\rightarrow H^{+}\bar{t}+X)_{net}=\nonumber\\
& \sigma (g\bar{b}%
\rightarrow H^{+}\bar{t})+\sigma (gg\rightarrow H^{+}\bar{t}b)  \nonumber \\
& -\sigma (g\rightarrow b\bar{b}\otimes g\bar{b}\rightarrow
H^{+}\bar{t}). \label{netgb}
\end{eqnarray}
Furthermore, for the study of the various differential
distributions one has to properly combine the $2\rightarrow2$ and
$2\rightarrow3$ processes mentioned above in order to reproduce
not only the total cross-section but also the \textit{correct
event kinematics.} The point is that we know the \textit{total
amount} of double counting but not a priori which part of this
value should be
subtracted from the $H^{+}\bar{t}$ process and which part from the $H^{+}%
\bar{t}b$ one. We apply here the method proposed in~\cite{TW} for
the analogous process of the single top quark production.
According to this method we use the cut on the transverse momenta
of the $b$-quark associated
with charged Higgs boson production to separate and recombine $gg$- and $\gb$%
-initiated processes.

\section{Signal versus background}
Concerning the signature, we consider the dominant
$t\bar{t}b\bar{b}$ final state for the combined signal
process~(\ref{tbh}). We focus here on the triple $b$-tagging case
and consider the situation where one top decays hadronically and
the other leptonically (including
 \textit{only} electron and muon decay {channels})
in order to reduce the combinatorics when both top quarks are
reconstructed. The branching
ratio of $t\bar{t}b\bar{b}\rightarrow b\bar{b}b\bar{b}\ell^{\pm}\nu q\bar{%
q^{\prime}}$ is $2/9\times2/3\times2=8/27$.

In order to decide whether a charged Higgs boson cross-section
leads to a detectable signal, we have to compute the background
rate.  The main QCD backgrounds leading to the same $t%
\bar{t}b\bar{b} $ signature and their respective cross-sections
are shown in Table \ref{bkg}.
\begin{table}[tbp]
\vskip -0.15cm
\begin{tabular}{llll}
& $\sigma(qq\rightarrow t\bar{t}b\bar{b}) $ & 6.62 fb & 0.266 pb \\
& $\sigma(gg\rightarrow t\bar{t}b\bar{b}) $ & 0.676 fb & 6.00 pb \\
& $\sigma(gb\rightarrow t\bar{t}b) $ & 1.22 fb & 4.33 pb \\
& Subtr. term & 0.72 fb & 2.1 pb \\
\end{tabular}
\vspace{0.2cm} \caption{The main background processes to the
signal (\ref{tbh}) at the Tevatron (2nd column) and the LHC (3rd
column) under cuts explained in the text. The corresponding
subtraction term is also shown. }\label{bkg}
\end{table}
For reconstructing the $t\bar{t}b(\bar{b}) $ final state, the
following steps are followed:
i) We reconstruct the W-boson mass from lepton and neutrino momenta: $%
M_{W1}^{rec}=(p_{\ell}+p_{\nu})^{2} $. The basic cuts for the
lepton (electron or muon) has been chosen as follows:
\begin{equation}
p_{T}^{\ell}>15\GeV,\, \, \, \, \, \, \, \, |\eta_{\ell}|<2.5,\,
\, \, \, p_{T}^{miss}>15\GeV\,\,; \label{leptoncut}
\end{equation}
ii) We reconstruct the mass of the second W-boson ($M_{W2}^{rec}
$). The following basic cuts for the jets were chosen for the
Tevatron (LHC):
\begin{equation}
p_{T}^{j,b}>20\, \, (30)\GeV\,, \, \, \ \ |\eta_{j}|<3,\, \,
|\eta_{b}|<2\, . \label{jetcut}
\end{equation}
Then we form and keep all $m_{t_{1}}=M_{W1b} $ and $%
m_{t_{2}}=M_{W2b} $ combinations for the first and second
top-quarks; iii) In the final step we form the $\chi$ function
\begin{eqnarray}
&\chi^2=(M_{W1}^{rec}-M_{W})^{2}+(M_{W2}^{rec}-M_{W})^{2}\nonumber\\
&+(m_{t_{1}}-\mt%
)^{2}+(m_{t_{2}}-\mt)^{2} \label{chidef}
\end{eqnarray}
for all combinations of $b $-jets, jets, lepton and neutrino and
choose the combination giving the smallest (best) value of the
$\chi$ function. After the reconstruction of the
$t\bar{t}b\bar{b} $ state one should reconstruct the charged
Higgs boson mass for the signal and the continuous $tb $
mass for the background. We assume that the  $b $-jet with the highest $%
p_{T} $ in $\bar{t}bt(\bar{b}) $ signature comes from the $H^{+}
$ decay. After all cuts are set up we obtained the signal and
background efficiencies, $S/B$ ratio and signal significance
$S/\sqrt{B}$-- see Tables 4,5 in the second paper of \cite{bggs}
for details. They will be used in the next section after
including the SUSY corrections.

For the calculation we use $\mt=175\GeV$, $m_{b}=4.6%
\GeV$ and the CTEQ4L set of PDFs\,\cite{PDFS1}. Here $ m_{t},m_b$
refer to the quark pole masses. The central value of the (common)
factorization and renormalization scale, $\mu_R$, for the signal
processes has been chosen equal to $\mH$, whereas that of the
background processes to $2\mt$.  We have checked  the uncertainty
of the signal due to variations of $\mu_R$ in the interval
$\mH/2<\mu_R<2 \mH$. We find that individual sub-channels show a
stronger dependence than the total cross-section, which varies
$\delta\sigma\sim 28\%$ at the Tevatron II and $\delta\sigma\sim
18\%$ at the LHC. We have also compared our CTEQ4 results with
the ones obtained with the MRST (central gluon)
PDFs\,\cite{PDFS2}. Again, significant deviations appear for some
of the individual sub-channels (up to $\sim 50\%$), but they are
compensated in the sum, leaving a $5-10\%$ thanks to the
substraction procedure in (\ref{netgb}).

\section{Supersymmetric quantum effects}

\begin{table}[tbp]
\begin{tabular}{cccccccc} &
$\mu$ &
$M_2$ & $m_{\tilde g}$ & $\mst1$ & $\msb1$ & $A_t$ & $A_b$ \\
\hline  A & \small-1000 & \small 200 & \small 1000 & \small 1000 &
\small 1000 & \small 500 & \small 500
\\ \hline  B &  \small -200 & \small 200 & \small 1000 & \small 500 & \small 500 & \small 500 & \small 500
\\ \hline C & \small 200 & \small 200 & \small 1000 & \small 500 & \small 500 & \small -500 & \small 500
\\ \hline D & \small 1000 & \small 200 & \small 1000 & \small 1000 & \small 1000 & \small -500 &
\small 500 \\
\end{tabular}
\vspace{0.2cm} \caption{Sets of MSSM parameters used in the
computation of the SUSY corrections to process (\ref{tbh}). All
masses and trilinear couplings in~GeV.}\label{tab:SUSYpar}
\end{table}

\begin{figure}
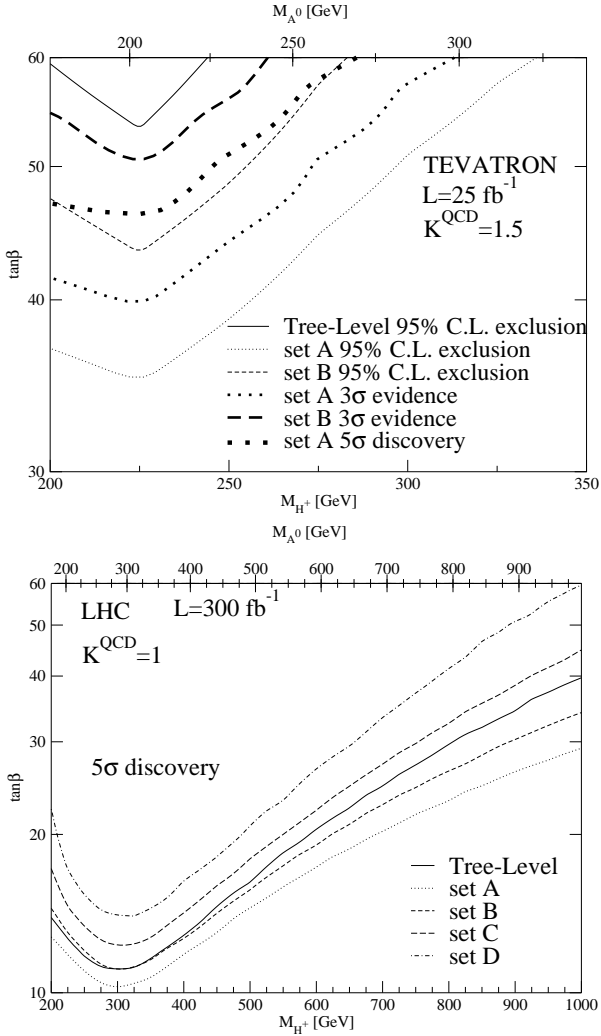

\begin{tabular}{c}
\resizebox{!}{6.8cm}{\includegraphics{TEV_discovery_mHc_ma_new.eps}}\\
\resizebox{!}{6.8cm}{\includegraphics*{LHC_discovery_mHc_ma_new.eps}}
\end{tabular}
\caption{%
Discovery regions for the Tevatron and the LHC, in the $\tb-\mH$
plane, at two levels of significance $S/\sqrt{B}=3,5\,$ for the
Tevatron, and $S/\sqrt{B}=5$ for the LHC.  The exclusion limit at
$95\%\,C.L.$ is also given in each case. We include the value of
$M_{A^0}$ corresponding to each $\mH$ according to the one-loop
formulae for the MSSM Higgs sector. The tree-level prediction can
be compared with the SUSY results for the various parameter sets
of Table \ref{tab:SUSYpar}. \label{fig:discovery} }
\end{figure}


In Figs.~\ref{fig:discovery}a, b we find the regions of the
$\tb-\mH$ plane in which the Tevatron and the LHC can find (or
exclude) the existence of the charged Higgs boson. The curves
correspond to the various sets of MSSM parameters indicated in
Table \ref{tab:SUSYpar}. For the Tevatron we have included a QCD
$K$-factor of $1.5$, whereas for the LHC we have just set $K=1$.
From these figures it is obvious that the presence of the SUSY
corrections alters significantly the  Higgs boson discovery
potential of the hadron colliders. It is also patent the Tevatron
could have a chance to find a charged Higgs boson in the
intermediate range $M_{H^{\pm}}<280\GeV$.

Let us now sketch how the various quantum corrections have been
computed. Among the plethora of possible corrections we
disregarded virtual supersymmetric effects on the $gqq$ and $ggg$
vertices and on the gluon propagators. We expect those to be of
order $(\aS/4\pi)\cdot(s/M_{SUSY}^{2})$ and thus suppressed by a
non-enhanced (i.e. \tb-independent) MSSM form factor coming from
the loop integrals. Therefore, we can neglect these contributions
as we are only considering effects of the form
$(\aS/4\pi)^n\cdot\tan^n\beta$ at large \tb\,. The cross-section
for the signal increases steeply with \tb\ and becomes highly
significant for \tb $>30$, while it is much smaller for \tb\ in
the low interval $2-20$ where the remaining SUSY corrections are
of the same order or even dominant, so our approximation is well
justified. Similarly, we neglect  all those electroweak
corrections in vertices and self-energies which are proportional
to pure $SU(2)_L\times U(1)_Y$ gauge couplings; in particular,
vertices involving electroweak gauge bosons and those involving
electroweak gauginos. Furthermore, we have checked that vertices
involving Higgs bosons exchange yield a very tiny overall
contribution, due to automatic cancellations arranged by the
underlying supersymmetry. Finally, there are the strong
gluino-squark diagrams and the $\tan\beta$-enhanced
higgsino-squark vertices implicit in chargino-neutralino loops.
We have extracted the parts of these interactions which are (by
far) the more relevant ones at high $\tan\beta$ and confirmed
that the remaining contributions are negligible.  In practice
this means that we may concentrate our analysis on the interval
$\tan\beta>20$ where we can be sure that our approximation does
include the bulk of the MSSM corrections while at the same time
the  cross-section of process (\ref{tbh}) starts to be
sufficiently large to consider it as an efficient mechanism for
charged Higgs boson production.  The leading contributions are
similar to those found in Ref. \,\cite{SUSYtbH} for the decay
process $t\rightarrow H^+\,b$, where we also refer for detailed
renormalization issues. Recently these dominant effects have been
conveniently described through an effective Lagrangian approach
that contains effective couplings absorbing both the leading SUSY
contributions and the known part of the QCD
corrections~\cite{eff}. At high $\tan\beta$ the most relevant
piece is the effective $tbH^{+}$-coupling as it carries the
leading part of the (appropriately resummed) quantum effects:
\begin{equation}
{\cal L}=\frac{gV_{tb}}{\sqrt{2\,}M_{W}}\,\frac{\overline
{m}_{b}(\mu_{R})\,\tan\beta}{1+\Delta m_{b}}\,H^{+}\overline{t}%
_{L}\,b_{R}+h.c.  \label{effecLag}
\end{equation}
The quantity $\Dmb$ above-- see (\cite{SUSYtbH,eff})-- contains
the bulk of the supersymmetric contributions. Although the
leading corrections have been identified, we have computed the
full set of one-loop SUSY diagrams for the relevant $tbH^{+}$
vertex, which involve similar bunch of diagrams as in the on-shell
case\,\cite{SUSYtbH}, but here we have got to account for the
off-shell external lines, which is a non-trivial task. The result
is that these off-shell corrections amount to a few percent
contribution that we have included in our numerical analysis.

Coming back to Figs.~\ref{fig:discovery} and Table
\ref{tab:SUSYpar}, let us remark that parameter set B constitutes
a typical case for moderately positive corrections. On the other
hand set A is over-optimized, and sets C and D give more and more
negative corrections. In all cases these corrections decouple
very slowly with the gluino mass, for fixed values of the other
masses. Furthermore, they exhibit a non-decoupling behavior when
\textit{all} SUSY parameters are scaled up by a common
factor\,\cite{SUSYtbH,eff}. This SUSY non-decoupling
behavior\,\cite{Haber} is associated to the properties of the
quantity $\Dmb$ entering eq.(\ref{effecLag}). The violation of
the decoupling theorem\,\cite{AC} is caused by the dimensionful
Higgs-quark-squark coupling, which increases arbitrarily with the
SUSY scale. Through this coupling the MSSM Higgs doublet that
interacts only with the top quark at the tree-level, effectively
couples to the bottom quark at one-loop. Therefore, that
interaction cannot be re-absorbed in the parameters of the
low-energy Higgs sector after integrating out the sparticle
masses.  This welcome feature is at the basis of the large
radiative corrections found in our study of the charged Higgs
production process (\ref{tbh}), and it could be responsible for
the eventual finding of this scalar boson. These facts were amply
exploited in Ref.\,\cite{strong} where the suggestion of the
present calculation was first made and the importance of its
correlation with the neutral Higgs production processes $p\overline{p}(pp)\rightarrow h\,%
\overline{b}b+X (h=h^{0},H^{0},A^{0})$ was first emphasized. Such
effects and correlations could eventually lead to the discovery
of SUSY.


Acknowledgments:  This collaboration is part of the network
``Physics at Colliders'' of the European Union under contract
HPRN-CT-2000-00149. The work of J.G. has been partially supported
by the European Union under contract No. HPMF-CT-1999-00150. The
work of J.S. has been supported in part by MECYT and FEDER under
project FPA2001-3598.

\providecommand{\href}[2]{#2}


\begin{thebibliography}{9}
\footnotesize

\bibitem{Haber} H. Haber, these proceedings.

\bibitem{HiggsRunII} M.~Carena \textit{et al.}, ``Report of the
Tevatron Higgs working group of the Tevatron Run 2 SUSY/Higgs
Workshop'', hep-ph/0010338, and references therein.

\bibitem{Hppt}
F.~Borzumati, J.~Kneur, N.~Polonsky, Phys.\ Rev.\ \textbf{D60}
(1999) 115011; D.J.~Miller \textit{et al.}, Phys.\ Rev.\
\textbf{D61} (2000) 055011.


\bibitem{Coarasa}
J.A.~Coarasa, J.~Guasch, J.~Sol\`{a}, \textit{Top quark and
charged Higgs at the Tevatron Run II}, hep-ph/9909397. Contributed
to Physics at Run II: Workshop on SUSY/Higgs, Fermilab, 1998, see
Ref.\,\cite{HiggsRunII}.

\bibitem{bggs}
A.~Belyaev, D.~Garcia, J.~Guasch, J.~Sol\`a, Phys.\ Rev.\ {\bf
D65} (2002) 031701(R) and JHEP 0206:059 (2002).

\bibitem{Reina} W.~Beenakker \textit{et
al.} Phys.\ Rev.\ Lett.\  {\bf 87} (2001) 201805; L.~Reina,
S.~Dawson, Phys.\ Rev.\ Lett.\  {\bf 87} (2001) 201804; see also
L. Reina and S. Dittmaier, these proceedings


\bibitem{shouhua}
S.-H. Zhu, hep-ph/0112109; T. Plehn, hep-ph/0206121.

\bibitem{Barnett}
R.M.~Barnett, H.E.~Haber, D.E.~Soper, Nucl.\ Phys.\  \textbf{B306}
(1988) 697; F.I.~Olness, W.~Tung, Nucl.\ Phys.\  \textbf{B308}
(1988) 813.

\bibitem{TW}
A.S.~Belyaev, E.E.~Boos, L.V.~Dudko, Phys.\ Rev.\  \textbf{D59}
(1999) 075001.

\bibitem{PDFS1}
H.L.~Lai \textit{et al.}, Phys.\ Rev.\  \textbf{D51} (1995) 4763.

\bibitem{PDFS2}
A.~Martin \textit{et al.}, Eur.\ Phys.\ J.\ \textbf{C4} (1998)
463.

\bibitem{SUSYtbH}
J.A.~Coarasa \textit{et al.}, Eur.\ Phys.\ J.\ \textbf{C2} (1998)
373; J. Guasch, R.A. Jim\'enez, J. Sol\`a,\, Phys. Lett. {\bf
B360} (1995) 47.

\bibitem{eff}
M.~Carena \textit{et al.},  Nucl.\ Phys.\ \textbf{B577} (2000) 88,
and references therein.

\bibitem{AC} T. Appelquist, J. Carazzone, {Phys. Rev.}
\textbf{D11} (1975) 2856.

\bibitem{strong} R.A. Jimenez, J. Sol\`a, Phys. Lett.B 389 (1996)
(53); J.A. Coarasa, R.A. Jimenez, J. Sol\`a,\, Phys. Lett. B 389
(1996) 312.

\end{thebibliography}
\end{document}